\documentclass[twocolumn,showpacs,preprintnumbers,amsmath,amssymb]{revtex4}

\def\undersim#1{\setbox9\hbox{${#1}$}{#1}\kern-\wd9\lower
    2.5pt \hbox{\lower\dp9\hbox to \wd9{\hss $_\sim$\hss}}}
\usepackage{amssymb}
\usepackage{amsmath}
\usepackage{graphicx}
\usepackage{color}

\def\undersim#1{\setbox9\hbox{${#1}$}{#1}\kern-\wd9\lower
    2.5pt \hbox{\lower\dp9\hbox to \wd9{\hss $_\sim$\hss}}}

\def\mk{{\mathbf p}}

\begin{document}

\title{Squeezed back-to-back correlation of \boldmath{$D^0{\bar D}^0$} in relativistic
heavy-ion collisions \footnote{Supported by the National Natural Science Foundation
of China under Grant Nos. 11675034, 11647166, and 11675033.}}

\author{Ai-Geng Yang$^1$}
\author{Yong Zhang$^2$}
\author{Luan Cheng$^1$}
\author{Hao Sun$^1$\footnote{haosun@dlut.edu.cn}}
\author{Wei-Ning Zhang$^{1,3}$\footnote{wnzhang@dlut.edu.cn}}
\affiliation{$^1$School of Physics, Dalian University of Technology, Dalian,
Liaoning 116024, China\\
$^2$School of Mathematics and Physics, Jiangsu University of Technology, Changzhou,
Jiangsu 213001, China\\
$^3$Department of Physics, Harbin Institute of Technology, Harbin, Heilongjiang
150006, China\\}

%\date{\today}

\begin{abstract}
We investigate the squeezed back-to-back correlation (BBC) of $D^0\!{\bar D}^0$
in relativistic heavy-ion collisions, using the in-medium mass modification
calculated with a self-energy in hot pion gas and the source space-time
distributions provided by the viscous hydrodynamic code VISH2+1.  It is found 
that the squeezed BBC of $D^0\!{\bar D}^0$ is significant in peripheral Au+Au
collisions at the RHIC energy.  A possible way to detect the squeezed BBC 
in experiment is presented.
\end{abstract}

\maketitle

In high-energy heavy-ion collisions, the interactions of bosons with hadronic medium
before the thermal freeze-out of hadrons may cause the in-medium mass modification of
bosons and then lead to a squeezed back-to-back correlation (BBC) between detected boson
and anti-boson \cite{AsaCso96,AsaCsoGyu99,Padula06,Zhang15a}.  This BBC is related to the
Bogoliubov transformation between the creation (annihilation) operators of the quasiparticle
in medium  and the detected particle \cite{AsaCso96,AsaCsoGyu99,Padula06,Zhang15a}, and 
expected to be strong for the bosons with large mass under the same mass modification \cite{Zhang-EPJC16,Padula-JPG10}.  Thus the squeezed BBC is likely an observable for heavy 
mesons to investigate the interactions of particles with the hadronic medium.

Recently, $D$ mesons are measured in heavy-ion collisions at the Relativistic Heavy Ion
Collider (RHIC) by the STAR collaboration \cite{STAR-PRL14,STAR-NPA16} and at the Large
Hadron Collider (LHC) by the ALICE collaboration \cite{ALICE-PRL13,ALICE-PRC14,ALICE-JHEP16},
respectively.  Because containing a heavy quark (charm quark), which is believed to
experience the whole evolution of the quark-gluon plasma (QGP) created in relativistic
heavy-ion collisions, the analyses of experimental data of $D$ mesons are of great interest.
It is expected that the BBC of heavy quarks produced in the early stage of relativistic 
heavy-ion collisions may lead to a BBC of $D^0\!{\bar D}^0$ \cite{XLZhu-PLB07}.  
Because of the interactions of heavy quarks with the high density QGP and the violent 
expansion of QGP, the BBC of $D^0\!{\bar D}^0$ may exhibit different behaviors \cite{XLZhu-PLB07,ZhuXuZhuang-PRL08}.  However, it should be emphasized that not only the
interactions of heavy quarks with QGP but also the effects of hadronic medium on the detected
$D$ meson should be considered seriously in the correlation analyses.

In Ref. \cite{FMFK-PRC06}, C.~Fuchs, B.~Martemyanov, A.~Faessler, and M.~Krivoruchenko
calculate the in-medium self-energies of $D$ mesons at rest in a hot pion gas, induced
by resonance interactions with pions.  Considering that the hadronic medium is baryon 
dilute and meson rich in heavy-ion collisions at the RHIC and the LHC, in this study we 
calculate the in-medium mass shift and width of $D$ meson moving in the hot pion gas,
in the FMFK framework developed in Refs. \cite{FMFK-PRC06,MFFK-PRL04}.

In the FMFK framework, the self-energy $\Sigma$ of a $D$ meson in the pion gas can be
written as \cite{FMFK-PRC06}
\begin{equation}
\Sigma=-\int A_{\frac{1}{2}}\,\frac{1}{16\pi^3E_\pi}\,\frac{d^3p_\pi}{e^{E_\pi/T}-1},
\label{S2}
\end{equation}
for considering only the resonances with $\frac{1}{2}$ isospin.  Here, the forward
resonance amplitude $A_{\frac{1}{2}}$ can be written in the form
\cite{FMFK-PRC06}:
\begin{equation}
A_{\frac{1}{2}}=\sum_{j=0,1,2}\!\frac{8\pi\sqrt{s}}{k}\,\frac{(2j+1)}{(2j_1+1)(2j_2+1)}
\,\frac{-\sqrt{s}\Gamma_j^{D\pi}}{s-M_j^2+i\sqrt{s}\Gamma_j^{\rm tot}},
\label{A12}
\end{equation}
where $j=0,1,2$ correspond to the $s$-, $p$-, and $d$-wave resonances $D^*_0$,
$D^*$, and $D^*_2$ in the $D\pi$ system with masses $M_j$, partial and total widths
$\Gamma_j^{D\pi}$ and $\Gamma_j^{\rm tot}$, respectively; $j_1 = j_2 = 0$ are the
spins of the $D$ and the pion, respectively, and $k$ is the c.m. momentum.  The energy
dependence of the widths is given by \cite{FMFK-PRC06}
\begin{equation}
\Gamma_j^{D\pi}=\bigg(\frac{k}{k_0}\bigg)^{2j+1}\frac{M_j^2}{s}\Gamma_{j0}^{D\pi},
\label{G2}
\end{equation}
where $k_0$ is the c.m. momentum when the resonance energy is its mass, and
$\Gamma_{j0}^{D\pi}$ is the on-mass-shell decay width.  We list in Table I the values
of $M_j$ and $\Gamma_{j0}^{D\pi}$ for the resonances $D^*_0$, $D^*$, and $D^*_2$
\cite{FMFK-PRC06}.  With Eqs. (\ref{S2})~--~(\ref{G2}) and the data in Table I, one
can calculate the self-energy $\Sigma$ of $D$ meson in the hot pion gas.

\begin{table}[htbp]
\caption{Excited $D$-meson states which are taken into account as
resonances in $D\pi$ system \cite{FMFK-PRC06}. }
\label{labl2}
\begin{center}
\begin{tabular}{ccc}
\hline
~Resonance~~~~&~~~~$M_j$ (MeV/$c^2$)~~~~&~~~~$\Gamma^{D\pi}_{j0}$ (MeV/$c^2$)~\\
\hline
$D^*$ & 2008.5 & $\approx$ 0.1 \\
$D^*_0$ &$ 2308\pm 60$ & $276\pm 99$ \\
$D^*_2$ &$2461.6\pm 5.9$  &$ 45.6\pm 12.5$\\
\hline
\end{tabular}
\end{center}
\end{table}

\begin{figure}[htbp]
\includegraphics[scale=0.55]{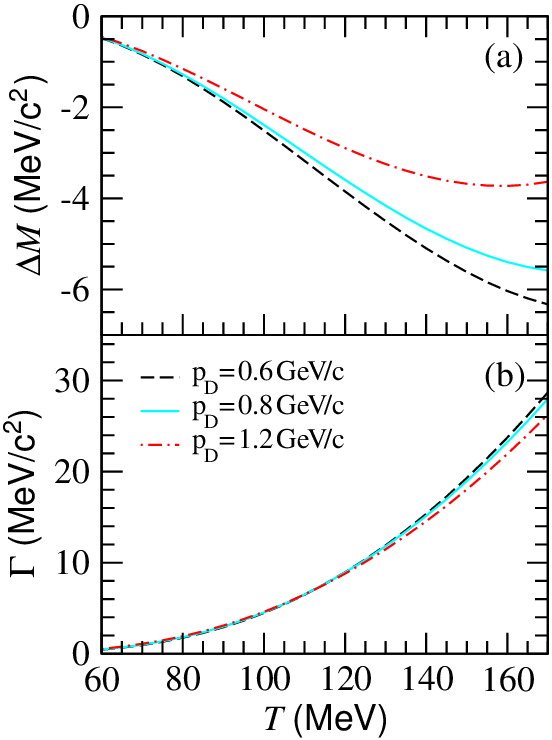}
\caption{(Color online) Mass shift $\Delta M=({\rm Re}\Sigma/2M_{\!D})$ and
width $\Gamma=(-{\rm Im} \Sigma/E_{\!D})$ of $D$ meson in the hot pion gas. }
\label{zFig-Dmw}
\end{figure}

We show in Fig.\,\ref{zFig-Dmw} the mass shift $\Delta M=({\rm Re}\Sigma/2M_{\!D})$
and width $\Gamma=(-{\rm Im} \Sigma/E_{\!D})$ of $D$ meson in the hot pion gas.
Here, $E_{\!D}=\sqrt{p_{\!_D}^2+M_{\!D}^2}$ and $M_{\!D}$ is taken to be
1.865\,GeV/$c^2$.  One can see that the mass-shift magnitude and the width increase
with temperature.  The mass-shift magnitude becomes smaller for larger $D$-meson
momentum $p_{\!_D}$ and the width decreases slightly with increasing $p_{\!_D}$.

The squeezed correlation function of boson-antiboson with momenta $\mk_1$  and $\mk_2$
is defined as \cite{AsaCsoGyu99,Padula06}
\begin{equation}
\label{BBCf}
C(\mk_1,\mk_2) = 1 + \frac{|G_s(\mk_1,\mk_2)|^2}{G_c(\mk_1,\mk_1) G_c(\mk_2,
\mk_2)},
\end{equation}
where $G_c(\mk_1,\mk_2)$ and $G_s(\mk_1,\mk_2)$ are the so-called chaotic and
squeezed amplitudes, respectively \cite{AsaCsoGyu99,Padula06}.
In a homogeneous medium with volume $V$ and
temperature $T$, the amplitudes $G_c(\mk_1,\mk_2)\propto V\delta_{\mk_1,\mk_2}$,
$G_s(\mk_1,\mk_2)\propto V\delta_{\mk_1,-\mk_2}$, and the squeezed correlation
function can be written as \cite{AsaCsoGyu99,Padula06,Zhang15a,Zhang-EPJC16}
\begin{eqnarray}
\label{BBCf1}
C(\mk,-\mk)&=&1+\frac{V^2|c_{\mk}\,s_{\mk}^*\,n_{\mk} +c_{-\mk}\,s_{-\mk}^*
\,(n_{-\mk}+1)|^2}{V^2[n_1(\mk)\,n_1(-\mk)]}\nonumber\\
&\equiv&1+B(\mk),
\end{eqnarray}
where $c_\mk$ and $s_\mk$ are the coefficients of Bogoliubov transformation
between the creation (annihilation) operators of the quasiparticle in medium
with modified mass $m_*$ and the free observed particle, $n_\mk$ is boson
distribution of quasiparticle, and $n_1(\mk)=|c_{\mk}|^2\,n_{\mk}+|s_{-\mk}|^2
(n_{-\mk}+1)$.  The function $B(\mk)$ is not zero only when $\mk_2=-\mk_1$, for
the homogeneous source.  Thus the squeezed correlation function is also called
the BBC function, which will be 1 if there is no in-medium mass modification
\cite{AsaCso96,AsaCsoGyu99,Padula06,Zhang15a,Zhang-EPJC16}.

For a finite source, there are still the squeezed correlations when $\mk_2\ne
-\mk_1$ but $\mk_2\approx -\mk_1$.  Considering a static source for simplicity
with the same temperature and Gaussian spatial distribution with standard deviation
$R$ \cite{AsaCsoGyu99,Padula06,Zhang-CPC15}, for $|\mk_1|=|\mk_2|=|\mk|$ we have
%\begin{eqnarray}
%\label{Gc}
%G_c(\mk_1,\mk_2)= \frac{E_p}{(2 \pi)^3 } \!\!\!\int \!\! d^3 r\,e^{i(\mk_1
%-\mk_2)\cdot \mr}\, e^{-\mr^2\!/\!2R^2} n_1({\mk}),
%\end{eqnarray}
%\begin{eqnarray}
%\label{Gs}
%&&\hspace*{-8mm}G_s(\mk_1,\mk_2)= \frac{E_p}{(2 \pi)^3 } \!\!\!\int
%\!\! d^3 r\,e^{i(\mk_1+\mk_2)\cdot \mr} \, e^{-\mr^2\!/\!2R^2}\nonumber\\
%&&\hspace*{8mm}\times\! \Big\{c_{\mk}\,s^*_{\mk}\,n_{\mk}+c_{-\mk}\,s^*_{-\mk}
%\,[n_{-\mk}+1] \Big\},
%\end{eqnarray}
%and
\begin{equation}
C(\mk_1,\mk_2)\!=\!1\!+\!e^{-2\mk^2\!R^2[1+\cos(\alpha)]}B(\mk)\!\equiv\!1\!
+\!f(\alpha)B(\mk),
\end{equation}
where $\alpha$ is the angle between $\mk_1$ and $\mk_2$.

\begin{figure}[htbp]
\includegraphics[scale=0.50]{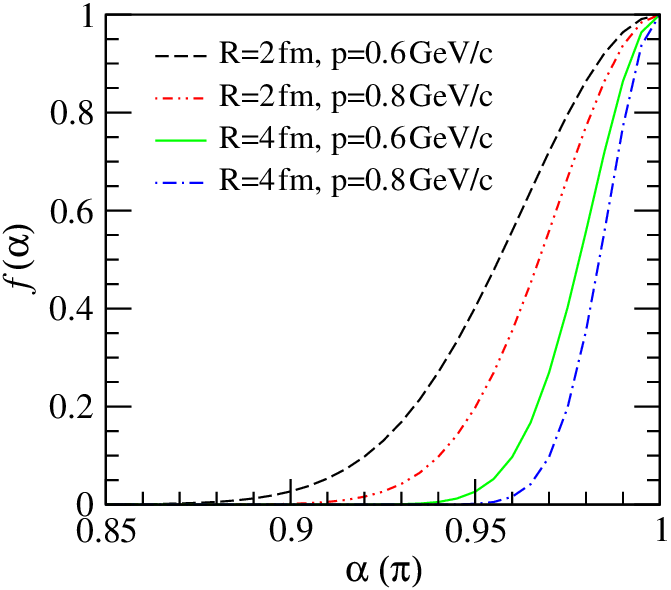}
\caption{(Color online) The function $f(\alpha)$ for different values of the source 
radius $R$ and particle momentum $p$. }
\label{zFf}
\end{figure}

We compare in Fig.\,\ref{zFf} the functions of $f(\alpha)$ for the different values
of source radius $R$ and particle momentum $p$.  Here, $f(\alpha)$ decreases with 
increasing $R$ and $p$ for fixed $\alpha$.  The squeezed correlation functions reach 
their maxima when $\mk_2=-\mk_1$.  There are still considerable squeezed correlations
when $\mk_2\ne -\mk_1$ but $\mk_2\approx -\mk_1$ for small sources.

For a general particle-emitting source evolving hydrodynamically, the chaotic and
squeezed amplitudes be expressed as
\cite{AsaCsoGyu99,Padula06,Zhang15a,Zhang-EPJC16,MakhSiny}
\begin{eqnarray}
\label{Gchydro}
&&\hspace*{-3mm}G_c({\mk_1},{\mk_2})\!=\!\int \frac{d^4\sigma_{\mu}(r)}{(2\pi)^3}
K^\mu_{1,2}\, e^{i\,q_{1,2}\cdot r}\,\! \Bigl\{|c'_{\mk'_1,\mk'_2}|^2\,
n'_{\mk'_1,\mk'_2}~~\nonumber \\
&& \hspace*{16mm}
+\,|s'_{-\mk'_1,-\mk'_2}|^2\,[\,n'_{-\mk'_1,-\mk'_2}+1]\Bigr\},\\[1ex]
\label{Gshydro}
&&\hspace*{-3mm}G_s({\mk_1},{\mk_2})\!=\!\int \frac{d^4\sigma_{\mu}(r)}{(2\pi)^3}
K^\mu_{1,2}\, e^{2 i\,K_{1,2}\cdot r}\!\Bigl\{s'^*_{-\mk'_1,\mk'_2}
c'_{\mk'_2,-\mk'_1}~~\nonumber \\
&& \hspace*{16mm}
\times n'_{-\mk'_1,\mk'_2}+c'_{\mk'_1,-\mk'_2} s'^*_{-\mk'_2,\mk'_1}
[n'_{\mk'_1,-\mk'_2} + 1] \Bigr\},\nonumber\\
\end{eqnarray}
where $d^4\sigma_{\mu}(r)$ is the four-dimension element of freeze-out
hypersurface, $q^{\mu}_{1,2}=p^{\mu}_1-p^{\mu}_2$, $K^{\mu}_{1,2}=(p^{\mu}_1
+p^{\mu}_2)/2$, and $\mk_i'$ is the local-frame momentum corresponding to $\mk_i
~(i=1,2)$.  In Eqs.\ (\ref{Gchydro}) and (\ref{Gshydro}), the quantities $c'_{\mk'_1,
\mk'_2}$ and $s'_{\mk'_1,\mk'_2}$ are the coefficients of Bogoliubov transformation
between the creation (annihilation) operators of the quasiparticles and the free
particles, and $n'_{\mk'_1,\mk'_2}$ is the boson distribution associated with the
particle pair \cite{AsaCsoGyu99,Padula06,Zhang15a}.
The squeezed BBC function obtained from Eqs. (\ref{BBCf}), (\ref{Gchydro}), and
(\ref{Gshydro}) is not only dependent on the mass modification, but also related
to the source space-time distribution and particle momentum
\cite{AsaCsoGyu99,Padula06,Zhang15a,Zhang-EPJC16}.

Relativistic hydrodynamics is an efficient tool for describing the evolution
of particle-emitting sources in high-energy heavy-ion collisions
\cite{Ris98,Kol03,Rom10,Gal13,Huo13,Sou16,Song-NST17}.
It has been widely used to study the momentum spectra
\cite{Gal13,Sou16,Pang16,Yang-CPC17}, anisotropic flows
\cite{Gal13,Huo13,Sou16,Song-NST17,Xu-CPL15,Pang16,Zhu-PRC17},
Hanbury--Brown--Twiss (HBT) radii \cite{Sou16,Efaaf-CPC12,Yang-CPC17,Pozek-PRC17},
and other observables of the final-state particles produced in relativistic
heavy-ion collisions.

\begin{figure}[htbp]
\includegraphics[scale=0.55]{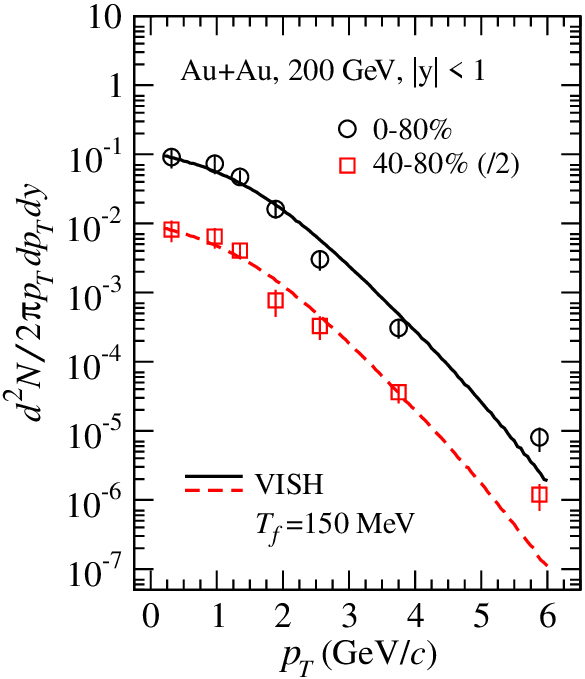}
\caption{(Color online) Transverse momentum spectra of $D^0$ meson calculated
with VISH2+1 for Au+Au at $\sqrt{s_{NN}}=200$\,GeV.  The experimental data are
obtained from the STAR Collaboration measurements \cite{STAR-PRL14}. }
\label{zFig-Dspe}
\end{figure}

We plot in Fig.\,\ref{zFig-Dspe} the transverse-momentum spectra of $D^0$
meson for Au+Au collisions at $\sqrt{s_{NN}}=200$\,GeV.  Here, the solid
and dashed lines are for the collisions in centrality intervals 0\%--80\%
and 40\%--80\%, respectively, calculated by the viscous hydrodynamic code
VISH2+1 \cite{VISH2+1} under the MC-Glb initial condition\cite{VISHb}.
The ratio of the shear viscosity to entropy density of the QGP is taken
to be 0.08 \cite{Shen11-prc,Qian16-prc} and the freeze-out temperature
is taken to be 150\,MeV.  The experimental data of the spectra are measured
by the STAR collaboration \cite{STAR-PRL14}.  One can see that the spectra
calculated by the VISH2+1 with $T_f=150$\,MeV are approximately consistent
with the data in $p_T<4$\,GeV/$c$.

\begin{figure}[htbp]
\includegraphics[scale=0.65]{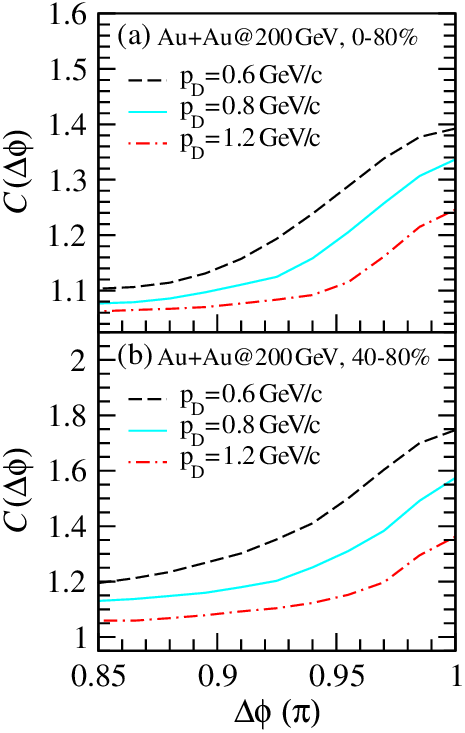}
\caption{(Color online) Squeezed BBC functions of $D^0\!{\bar D}^0$ for Au+Au 
collisions at $\sqrt{s_{NN}}=200$\,GeV. }
\label{zFig-DBBC}
\end{figure}

For the hydrodynamic sources for Au+Au collisions at $\sqrt{s_{NN}}=200$\,GeV,
we calculate the squeezed BBC functions $C(\Delta\phi)$ of $D^0\!{\bar D}^0$ 
as shown in Fig.\,\ref{zFig-DBBC}.  Here $\Delta\phi$ is the angle between the 
transverse momenta of $D^0$ and ${\bar D}^0$.  The dashed, solid, and dot-dashed 
curves represent the results calculated in the momentum regions 0.55--0.65\,GeV, 
0.75--0.85\,GeV, and 1.15--1.25\,GeV, respectively,
where the in-medium mass of $D^0$ meson, $m_*$, is taken according to the Breit-Wigner
distribution ${\cal D}(m_*,\Delta M,\Gamma)\!=\!(\Gamma/2\pi)/[(m_*\!-\!\Delta
M\!-\!M_{\!D})^2\!+\!(\Gamma/2)^2]$ with the $\Delta M$ and $\Gamma$ values when
$p_{\!_D}=$ 0.6, 0.8, and 1.2 GeV/$c$, respectively.  The more serious in-medium
mass modification at the lower $p_{\!_D}$ (see Fig.\,\ref{zFig-Dmw}) may lead
to a higher squeezed BBC than that at the higher momentum.  On the other hand, the 
more serious oscillations of single-event squeezed BBC functions at higher momentum
\cite{Zhang15a} may also lead to a lower squeezed BBC after averaged over events.
The squeezed BBC functions for centrality interval 40\%--80\% are higher than those 
for centrality interval 0\%--80\%.  The reason for this is mainly that the source
temporal distribution is narrow in peripheral collisions \cite{Zhang-EPJC16}.
The contributions to the squeezed BBC functions at lower $\Delta\phi$ are mainly from
the more peripheral collisions, which not only have the strong squeezed BBC but also
slow varying $f(\alpha)$ functions because of the smaller source size.

\begin{figure}[htbp]
\includegraphics[scale=0.65]{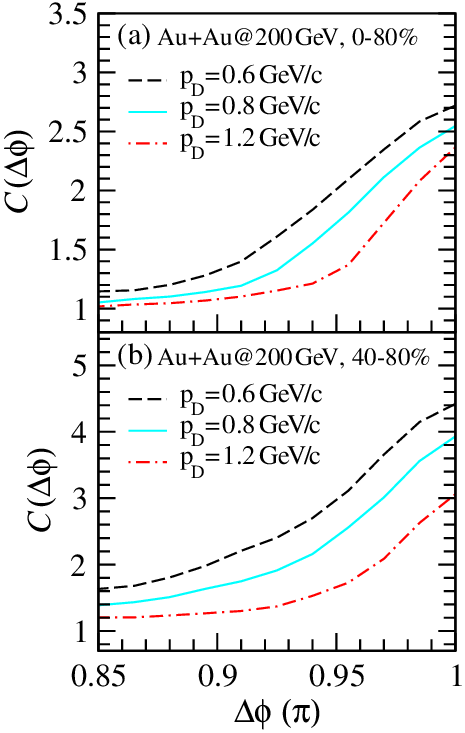}
\caption{(Color online) Squeezed BBC functions of $D^0\!{\bar D}^0$ for Au+Au 
collisions at $\sqrt{s_{NN}}=200$\,GeV, calculated with conditions $m_*<M_*$}
\label{zFig-DBBC-L}
\end{figure}

Considering that the in-medium mass shift is negative, the $D$ mesons with the
in-medium masses $m_*$ smaller than the peak of in-medium mass distribution,
$M_*$, have a large average mass difference to $M_{\!D}$, and thus have a
significant squeezed BBC \cite{YZ-WNZ16}.  We plot in Fig.\,\ref{zFig-DBBC-L} 
the squeezed BBC functions $C(\Delta \phi)$ of $D^0\!{\bar D}^0$ for the sources 
as in Fig.\,\ref{zFig-DBBC} but the in-medium masses $m_*$ of $D^0$ mesons are taken 
according to the Breit--Wigner distribution with $m_*<M_*$.  One can see that the 
squeezed BBC functions are more significant in this case than those in 
Fig.\,\ref{zFig-DBBC}.  The maxima of the squeezed BBC functions in 
Fig.\,\ref{zFig-DBBC-L} are about twice those in Fig.\,\ref{zFig-DBBC}.  Considering
that the number of $D^0$ mesons with $m_*<M_*$ is about half of the total number,
the errors will increase by about $\sqrt{2}$, which is smaller than 2.  Thus this
way provides a possibility to detect the squeezed BBC in experiment if the in-medium 
mass shift can be determined by experimental measurements.

In summary, we have investigated the squeezed BBC functions of $D^0\!{\bar D}^0$
in Au+Au collisions at $\sqrt{s_{NN}}=200$\,GeV and in centrality intervals 0\%--80\%
and 40\%--80\%.  The source evolution is described by viscous hydrodynamics, VISH$2+1$
code, and the mass modifications used are calculated in the FMFK framework.
It is found that the squeezed BBC of $D^0\!{\bar D}^0$ is observable in Au+Au collisions
at the RHIC energy, especially in peripheral collisions.  The squeezed BBC is obviously
greater than 1 when the momenta of the two mesons are approximately antiparalleled.
For a fixed angle between the momenta of the two $D$ mesons, the squeezed BBC decreases
with increasing momentum.  In experiment, researchers may use the $D$ mesons with the 
masses smaller than the maximum observed to detect the squeezed BBC and to analyze the 
mass modifications. On the other hand, the investigations of removing the background of 
heavy-quark correlations in squeezed BBC analyses will be of interest.


\begin{thebibliography}{99}
\bibitem{AsaCso96}
M. Asakawa and T. Cs\"org\H o, Heavy Ion Physics {\bf 4}, 233 (1996);
hep-ph/9612331.

\bibitem{AsaCsoGyu99}
M. Asakawa, T. Cs\"org\H o and M. Gyulassy, Phys. Rev. Lett. {\bf 83},
4013 (1999).

\bibitem{Padula06}
S. S. Padula, G. Krein, T. Cs\"org\H{o}, Y. Hama, P. K. Panda, Phys.
Rev. C {\bf 73}, 044906 (2006).

\bibitem{Zhang15a}
Y. Zhang, J. Yang, W. N. Zhang, Phys. Rev. C {\bf 92}, 024906 (2015).

\bibitem{Zhang-EPJC16}
Y. Zhang and W. N. Zhang, Eur. Phys. J. C {\bf 76}, 419 (2016).

\bibitem{Padula-JPG10}
S. S. Padula, D. M. Dudek, and O. Socolowski Jr, J. Phys. G {\bf 37}, 094056 (2010).

\bibitem{STAR-PRL14}
L. Adamczyk {\it et al.} (STAR Collaboration), Phys. Rev. Lett. {\bf 113},
142301 (2014).

\bibitem{STAR-NPA16}
M. Lomnitz for STAR Collaboration, Nucl. Phys. A {\bf 956}, 256 (2016).

\bibitem{ALICE-PRL13}
B. Abelev {\it et al.} (ALICE Collaboration), Phys. Rev. Lett. {\bf 111},
102301 (2013).

\bibitem{ALICE-PRC14}
B. Abelev {\it et al.} (ALICE Collaboration), Phys. Rev. C {\bf 90},
034904 (2014).

\bibitem{ALICE-JHEP16}
J. Adam {\it et al.} (ALICE Collaboration), JHEP {\bf 03}, 081 (2016).

\bibitem{XLZhu-PLB07}
X. Zhu, M. Bleicher, S. L. Huang, K. Schweda, H. St\"{o}cker, N. Xu, P. Zhuang,
Phys. Lett. B {\bf 467} 366 (2007).

\bibitem{ZhuXuZhuang-PRL08}
X. Zhu, N. Xu, P. Zhuang, Phys. Rev. Lett. {\bf 100} 152301 (2008).

\bibitem{FMFK-PRC06}
C. Fuchs, B. V. Martemyanov, A. Faessler, and M. I. Krivoruchenko, Phys. Rev.
C {\bf 73}, 035204 (2006).

\bibitem{MFFK-PRL04}
B. V. Martemyanov, A. Faessler, C. Fuchs, and M. I. Krivoruchenko, Phys. Rev.
Lett. {\bf 93}, 052301 (2004).

\bibitem{Zhang-CPC15}
Y. Zhang, J. Yang, and Wei-Ning Zhang, Chinese Physics C {\bf 39},
034103 (2015).

\bibitem{MakhSiny}
A. Makhlin and Yu. M. Sinyukov, Sov. J. Nucl. Phys. {\bf 46}, 354
(1987); Yu. M. Sinyukov, Nucl. Phys. {\bf A566}, 589c (1994).

\bibitem{Ris98}
D. H. Rischke, arXiv:nucl-th/9809044

\bibitem{Kol03}
P. F. Kolb and U. Heinz, arXiv:nucl-th/0305084

\bibitem{Rom10}
P. Romatschke, Int. J. Mod. Phys. E, {\bf 19}: 1 (2010)

\bibitem{Gal13}
C. Gale, S. Jeon, and B. Schenke, Int. J. of Mod. Phys. A,  {\bf 28}:
1340011 (2013)

\bibitem{Huo13}
P. Huovinen, Int. J. Mod. Phys. E, {\bf 22}: 1330029 (2013)

\bibitem{Sou16}
R. Derradi de Souza, T. Koide, T. Kodama, Prog. Part. Nucl. Phys.,
{\bf 86}: 35 (2016).

\bibitem{Song-NST17}
Huichao Song, You Zhou, K. Gajdo\u{s}ov\'a, Nucl. Sci. Tech. {\bf 28} 996 (2017).

\bibitem{Pang16}
Long-Gang Pang, Hannah Petersen, Guang-You Qin, Victor Roy, and Xin-Nian Wang,
Eur. Phys. J. A {\bf 52} 97 (2016).

\bibitem{Yang-CPC17}
Jing Yang, Wei-Ning Zhang, Yan-Yu Ren, Chinese Phys. C {\bf 41} 084102 (2017).

\bibitem{Xu-CPL15}
Jiechen Xu, Jinfeng Liao, Miklos Gyulassy, Chin. Phys. Lett. {\bf 32}, 092501
(2015).

\bibitem{Zhu-PRC17}
Xiangrong Zhu, You Zhou, Haojie Xu, Huichao Song, Phys. Rev. C {\bf 95}
044902 (2017).

\bibitem{Efaaf-CPC12}
Efaaf M. J., SU Zhong-Qian, ZHANG Wei-Ning, Chinese Physics C {\bf 36}
410 (2012).

\bibitem{Pozek-PRC17}
Piotr Bo\.{z}ek, Phys. Rev. C {\bf 95} 054909 (2017).

\bibitem{VISH2+1}
H. Song and U. Heinz, Phys. Lett. B {\bf 658}, 279 (2008);
H. Song and U. Heinz, Phys. Rev. C {\bf 77}, 064901 (2008).

\bibitem{VISHb}
C. Shen, Z. Qiu, H. Song, J. Bernhard, S. Bass, U. Heinz,
arXiv:1409.8164; https://u.osu.edu/vishnu/.

\bibitem{Shen11-prc}
C. Shen, U. Heinz, P. Huovinen, H. Song, Phys. Rev. C {\bf 84},
044903 (2011).

\bibitem{Qian16-prc}
J. Qian, U. Heinz, J. Liu, Phys. Rev. C {\bf 93}, 064901 (2016).

\bibitem{YZ-WNZ16} 
Y. Zhang and W. N. Zhang, arXiv:1611.05770[nucl-th]. 

\end{thebibliography}
\end{document}